\documentclass[%
 reprint,
nobibnotes,
 amsmath,amssymb,
 aps,
]{revtex4-1}

\usepackage{graphicx}% Include figure files
\usepackage{dcolumn}% Align table columns on decimal point
\usepackage{bm}% bold math
\usepackage[titletoc,title]{appendix}
\usepackage{amsmath}% http://ctan.org/pkg/amsmath
\begin{document}

%\preprint{APS/123-QED}

\title{Equilibrium cluster fluids: Pair interactions via inverse design}

\author{Ryan B. Jadrich}
%\email{rjadrich@utexas.edu}
\affiliation{McKetta Department of Chemical Engineering, University of Texas at Austin, Austin, Texas 78712, USA}

\author{Jonathan A. Bollinger}
%\email{jonathanabollinger@gmail.com}
\affiliation{McKetta Department of Chemical Engineering, University of Texas at Austin, Austin, Texas 78712, USA}

\author{Beth A. Lindquist}
%\email{kpj@che.utexas.edu}
\affiliation{McKetta Department of Chemical Engineering, University of Texas at Austin, Austin, Texas 78712, USA}

\author{Thomas M. Truskett}
%\email{truskett@che.utexas.edu}
\email{truskett@che.utexas.edu \dag Electronic Supplementary Information: three cluster fluid simulation movies (n8\textunderscore eta0.06\textunderscore movie, n16\textunderscore eta0.06\textunderscore movie and n32\textunderscore eta0.06\textunderscore movie) located at \emph{http://www.truskettgroup.com/papers/} corresponding to the snapshots in Fig.\ref{sch:FigureSNAPSHOTS} of the main text.}
\affiliation{McKetta Department of Chemical Engineering, University of Texas at Austin, Austin, Texas 78712, USA}

\date{\today}

\begin{abstract}

\noindent Inverse methods of statistical mechanics are becoming productive tools in the design of materials with specific microstructures or properties. While initial studies have focused on solid-state design targets (e.g, assembly of colloidal superlattices), one can alternatively design fluid states with desired morphologies. This work addresses the latter and demonstrates how a simple iterative Boltzmann inversion strategy can be used to determine the isotropic pair potential that reproduces the radial distribution function of a fluid of amorphous clusters with prescribed size. The inverse designed pair potential of this ``ideal'' cluster fluid, with its broad attractive well and narrow repulsive barrier at larger separations, is qualitatively different from the so-called SALR form most commonly associated with equilibrium cluster formation in colloids, which features short-range attractive (SA) and long-range repulsive (LR) contributions. These differences reflect alternative mechanisms for promoting cluster formation with an isotropic pair potential, and they in turn produce structured fluids with qualitatively different static and dynamic properties.  Specifically, equilibrium simulations show that the amorphous clusters resulting from the inverse designed potentials display more uniformity in size and shape, and they also show greater spatial and temporal resolution than those resulting from SALR interactions.

\end{abstract}

\pacs{Valid PACS appear here}% PACS, the Physics and Astronomy
                             % Classification Scheme.
%\keywords{Suggested keywords}%Use showkeys class option if keyword
                              %display desired
\maketitle

%\tableofcontents

\section{Introduction}

The computational design of interactions for targeted self assembly is a powerful approach in the search for new materials with specified microstructures, properties, or functionality. It is typically pursued via a strategy where the macroscopic behaviors of a subset of promising systems with different microscopic interactions (e.g., patchiness~\cite{doi:10.1021/nl0493500,doi:10.1021/la0513611,C0CP02296A,PhysRevLett.103.237801,doi:10.1021/jp404053t,C4SM01646G} or shape~\cite{C4SM01646G,Glotzer2004,Damasceno27072012}) are characterized by extensive ``forward'' molecular simulation calculations and compared to one another using appropriate figures of merit. Such forward approaches have been instrumental in discovering novel organizational motifs in crystalline or microphase-separated solids. However, since forward calculations (or experiments) can be expensive and time-consuming for complex systems, this method is perhaps most useful where physical intuition can guide the selection of the microscopic interactions to be considered. For example, possible locations and sizes of attractive ``patches'' on colloidal particles could be chosen \emph{a priori} by considering how these variables affect mutual patch alignment with nearest neighbors when particles are in a desired superlattice structure versus other competing morphologies~\cite{doi:10.1021/la0513611,B614955C,C0CP02296A,PhysRevLett.103.018101,Kraft03072012,C3SM52094C,Granick2011}.

For less intuitive materials design problems, systematic alternatives to forward searches may be helpful. Inverse methods of statistical mechanics, which formally optimize microscopic interactions toward attainment of a desired macroscopic outcome, are one such emerging class of complementary techniques~\cite{B814211B,AIC:AIC14491,doi:10.1021/mz400038b,C4SM01923G}. Inverse approaches have been recently applied to gain insights into nontrivial materials design problems including the search for isotropic, repulsive interactions that can stabilize low-coordinated crystals in two (e.g., honeycomb and square)~\cite{1.3576141,PhysRevX.4.031049,C0SM01205J} or three (e.g., diamond and simple cubic)~\cite{1.4790634,C3SM27785B,1.4825173,PhysRevX.4.031049} dimensions. However, such methods are not limited to designing solid-state targets, and standard tools could in principle be exploited to find interactions that imbue equilibrium \emph{fluid} states with desired microstructural features. Some of the organizational motifs of these designer fluids could, in turn, be captured in non-equilibrium solid states (e.g., gels or glasses formed from the flud via a rapid quench or compression).  

Here, we present a simple methodology for the inverse design of fluid structure via optimization of an isotropic pair interaction. It comprises two steps: generation of a configurational ensemble of target microstructures via simulations using an artificially complex, many-body interaction chosen to guarantee assembly of the desired morphology, and then use of a tool from systematic coarse-graining~\cite{noid2013,noid2011,shell2011,shell2012,doi:10.1021/ct900369w,doi:10.1021/ma0700983} to reduce the many-body interaction to an effective pair potential.  In the present work, we adopt iterative Boltzmann inversion (IBI) for the latter step, which uniquely determines the pair potential that will generate the radial distribution function (RDF) of the target ensemble at equilbrium. As a first application of this methodology, we attempt to inverse design a pairwise potential that forms a fluid of ``ideal'' amorphous equilibrium clusters of prescribed size. Clustered fluids of colloidal particles have attracted considerable interest due their novel multiscale structure, their rich dynamic and rheological properties, and their potential functionality ~\cite{GodfrinWagnerLiu2014,PhysRevE.91.042312,ManiBolhuis2014,ToledanoSciortino2009,Sciortino2004,LeoLue2014,Zhang2009,malins2011effect,malins2009geometric,Johnston2012,PorcarLiu2010,SorarufSchreiber2014,Stradner2004,Yearley2014,Charbonneau,Coniglio}. 

The classic paradigm for forming equilibrium clusters from an isotropic pair potential focuses on models that exhibit a combination of short-range attractive (SA) and longer-range repulsive (LR) contributions, commonly referred to as an SALR model~\cite{GodfrinWagnerLiu2014,PhysRevE.91.042312,ManiBolhuis2014,ToledanoSciortino2009,Sciortino2004,LeoLue2014,Zhang2009,malins2011effect,malins2009geometric,Charbonneau,Coniglio}.  Various functional forms for the attractions have been studied (modeling, e.g., polymer-mediated depletion forces between colloids), typically in combination with a repulsive Yukawa tail to model weakly screened Coulombic interactions. The attractions drive particle association, but the longer-ranged replusions lead to self-limited growth (i.e., finite sized aggregates). In contrast to systems lacking competitive repulsions, the formation of clusters in SALR fluids can either suppress macroscopic phase separation to lower temperature and density or eliminate it altogether~\cite{ArcherWilding,PhysRevE.91.042312,GodfrinWagnerLiu2014}. 

Although the SALR model qualitatively captures the effective pair potential and equilibrium cluster behavior seen in some types of experimental systems (e.g., mixtures of charged-stabilized colloids with weakly interacting polymers), it does not generate microstructures reflecting properties typically expected in the idealized picture~\cite{doi:10.1021/jp011646w} of such cluster phases~\cite{PhysRevE.91.042312,ManiBolhuis2014,ToledanoSciortino2009,Sciortino2004,LeoLue2014,Zhang2009,malins2011effect,malins2009geometric,Coniglio,ArcherWilding} that are found in other experimental systems~\cite{park2014terminal,xia2012self}, which we denote here as \emph{ideal cluster} (IC) fluids: particle assemblies that are monodisperse, spherical, long-lived, and fluid-like in terms of inter- and intra-cluster structure and mobility. Recently, Glotzer, Kotov, and coworkers have demonstated that a many-body potential which incorporates environment-dependent charge renormalization during assembly can lead to clusters that are monodisperse, spherical, and amorphous~\cite{Nguyen23062015,park2014terminal,xia2012self}. The role of surface charge renormalization has been studied by others as well~\cite{doi:10.1021/ct501067t,klix2013novel}. However, whether many-body interactions are in general neccessary to assemble the complex multiscale structures of IC fluids has been an open question. Additionally, keeping to the level of a pair interaction has the benefit that, through a systematic mapping, it can be regarded as a low dimensional approximation of a given many-body interaction. The simplified form can then yield key physical insights.

Here, we show that one can inverse design pair potentials that readily assemble into IC fluids under equilibrium conditions. Interestingly, these potentials exhibit a broad attractive well together with a narrow repulsive barrier at larger separations, which--while also a competitive balance between two interactions--is qualitatively different from those of the SALR fluid.  These differences imply distinct physics governing cluster formation in IC and SALR fluids, and we compare the static and dynamic properties of clusters in these systems, introducing a new metric for cluster lifetime to quantitatively characterize the latter. In the analysis, we also discuss practical aspects in the inverse design of pair potentials for complex fluids using IBI.   

The remainder of this paper is organized as follows. Section 2 outlines the constrained model used to generate configurational ensembles of the targeted ICs, the IBI inverse design method employed to discover the final IC potentials, the SALR model used and the metric we introduce for the cluster lifetime analysis. Results of the inverse cluster design and comparisons beween IC and SALR fluids are presented in Section 3. The paper is concluded in Section 4 with a discussion of future goals and possible improvements to the approach.

\section{Methods}

\subsection{Constrained Monte Carlo simulations}

The first step in the inverse design of ICs is to produce a \emph{physically realizable} target RDF, \(g_{\text{tgt}}(r)\), corresponding to a configurational ensemble with the desired structural properties. To generate such RDFs, we utilize constrained Monte Carlo simulations of \(N=2048\) total hard core (HC) particles of diameter \(d\), which are divided into equisized amorphous assemblies of either \(n_{\text{tgt}}=8\), 16, or 32 particles, each representing a single cluster. To enforce cluster association, single-particle translations are constrained by a many-body intra-cluster potential acting on the instantaneous cluster radius of gyration, \(R\), as
\begin{equation} \label{eqn:rg_potential}
\beta\varphi_{\text{intra}}(R)\equiv A (R^{2}-\overline{R}^2)^2
\end{equation}
where \(\beta=1/(k_{\text{B}}T)\) (\(k_{\text{B}}\) is Boltzmann's constant and \(T\) is temperature), \(A\) is a positive scalar amplitude, and \(\overline{R}\) is a target radius of gyration. For a given \(n_{\text{tgt}}\) there is a practical lower limit to what \(R\) values can be sampled by a cluster due to hard-core packing constraints; thus, any \(\overline{R}\) below this limit yields virtually identical behavior for appropriately chosen values of \(A\). For \(n_{\text{tgt}}=8\), 16, and 32, we use \(\overline{R}=0.6d\), \(0.8d\), and \(1.2d\) and \(A=300\), 265, and 170 respectively. For all particle packing fractions \(\eta=(\pi/6)Nd^{3}/V\) studied (where $V$ is volume), these parameters yield corresponding average radii of gyration, \(\langle R \rangle \approx 0.860d\), \(1.105d\), and \(1.476d\). The insensitivity of \(\langle R \rangle\) with respect to \(\eta\) for a given \(n_{\text{tgt}}\) is due to the preset compactness of the clusters.

We also introduce a longer-ranged Yukawa repulsion between the cluster center-of-mass (COM) pairs, which improves convergence of the IBI scheme (discussed below) and is defined by
\begin{equation} \label{eqn:com_potential}
\beta\varphi_{\text{COM}}(r_{\text{COM}})\equiv \dfrac{B}{r_{\text{COM}}} \textnormal{exp}[-r_{\text{COM}}/z]
\end{equation}
where \(r_{\text{COM}}\) is the pair COM distance between two clusters and \(B\) and \(z\) are the repulsive amplitude and range, respectively. We set \(z=0.12\) for all systems and \(B=1\times10^{12}\), \(1\times10^{15}\), and \(1\times10^{21}\) for \(n_{\text{tgt}}=8\), 16, and 32, respectively, for all \(\eta\). These parameters furnish very steep, hard-core-like repulsions around the clusters with effective hard-core diameters of \(d_{\text{eff}}\approx3.18d\), \(3.98d\), and \(5.60d\), respectively. From \(d_{\text{eff}}\), we can also obtain the effective volume fraction of whole clusters (treating them as renormalized objects) via the expression \(\eta_{\text{eff}}=\eta d_{\text{eff}}^3/(n_{\text{tgt}} d^3)\).

Given these definitions, we propagate the Monte Carlo trajectories via cluster COM translational moves (\(10\%\)) and single-particle displacements (\(90\%\)). Note that cluster rotational moves are unnecessary as the single-particle moves are sufficient to randomize the intra-cluster structures.

Once \(g_{\text{tgt}}(r)\) has been obtained, we must smooth out the discontinuous peak at contact that results from the use of the hard-core constraint, so as to be consistent with the use of continuous pair potentials as required by the molecular dynamics IBI framework. To do this, we construct an approximate hard-core mapping onto a steep, purely repulsive 50-25 Weeks-Chandler-Andersen (WCA) potential~\cite{HansenMcDonald2006} (in dimensionless form)
\begin{equation} \label{eqn:wca_potential}
\beta\varphi_{\text{WCA}}(r)\equiv H(r_{0}-r)\Bigg(4 \bigg[\bigg(\dfrac{d_{\text{WCA}}}{r}\bigg)^{50}-\bigg(\dfrac{d_{\text{WCA}}}{r}\bigg)^{25}\bigg]+1\Bigg)
\end{equation}
where \(H(x)\) is the Heaviside step function, \(d_{\text{WCA}}\) is the effective core diameter and \(r_{0}\equiv 2^{1/25}d_{\text{WCA}}\)). The mapping uses a linear extrapolation of the hardcore target RDF, \(g_{\text{tgt,HC}}(r)\), near contact into the core region to locally approximate the well known cavity distribution function, \(y_{\text{tgt,HC}}(r)\)~\cite{HansenMcDonald2006}. It is generally accepted that \(y_{\text{tgt,HC}}(r)\approx y_{\text{tgt,SC}}(r)\), where \(y_{\text{tgt,SC}}(r)\) is any steeply repulsive soft-core (SC) analog that approximates the hard core version via some non-unique mapping criterion. For our purposes, the simplest mapping, \(d_{\text{WCA}}=d\), is sufficient. The final soft-core profile is constructed as \(g_{\text{tgt,WCA}}(r)\approx \textnormal{exp}[-\beta \varphi_{\text{WCA}}(r)]y_{\text{tgt,HC}}(r)\), which we denote as \(g_{\text{tgt}}(r)\) in the remaining sections.

\subsection{Iterative Boltzmann inversion}

IBI is a conceptually simple and popular approach for solving the inverse statistical-mechanical problem of discovering the unique pair-potential \(u(r)\) corresponding to a particular RDF~\cite{noid2013,noid2011,shell2011,shell2012,doi:10.1021/ct900369w}. In general, there is no guarantee that such a potential exists according to the Henderson theorem~\cite{henderson1974uniqueness}; however, if the potential exists, %it is unique and % *** ``unique portion'' is redundant from previous sentence 
IBI is a suitable tool for recovering it. Inverse designed potentials depend on the state point of interest (\(\eta\) and \(T\) dependent); however, varying \(T\) at fixed \(\eta\) leads to trivial rescaling of the potential, thus all potentials are reported in units of thermal energy for generality. The explicit density dependence of our potentials is discussed in Section 3.2.

\begin{figure}
  \includegraphics{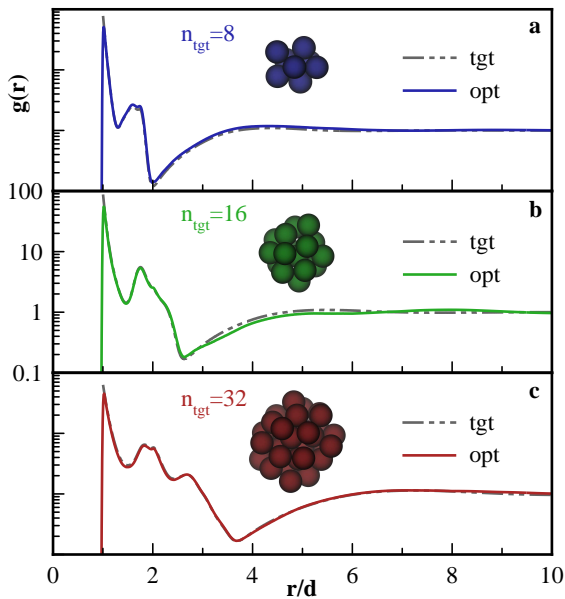}
  \caption{Comparison of the target (hard-core) and optimized radial distribution functions \(g(r)\) for \(n_{\text{tgt}}=8\), 16, and 32 at packing fraction \(\eta=0.04\).}
  \label{sch:FigureRDFs}
\end{figure}

The IBI procedure requires an initial-guess potential \(u_{1}(r)\), which at the lowest densities we take to be the target potential of mean force \(u_{1}(r)\equiv-k_{B}T\textnormal{ln}[g_{\text{tgt}}(r)]\). At higher densities, we use converged results from the lower densities. Simulation of \(u_{1}(r)\) provides the first trial RDF, \(g_{1}(r)\), and a new potential is calculated according to the general formula
\begin{equation} \label{eqn:ibi_equation}
u_{i+1}(r)\equiv u_{i}(r)+\alpha_{\text m} k_{B}T \textnormal{ln}\bigg[\dfrac{g_{i}(r)}{g_{\text{tgt}}(r)}\bigg]
\end{equation}
where \(\alpha_{\text m}\) is a mixing parameter to help control the convergence. The simulation step and potential update steps are carried out successively until satisfactory convergence in \(u(r)\) [\(g(r)\)] is achieved. For our highly structured RDFs (orders of magnitude variation; see Fig.~\ref{sch:FigureRDFs}), \(\alpha_{\text m}\) is best kept very small and \(\alpha_{\text m} \approx 0.005-0.02\) provides the best convergence while maintaining a quasi-equilibrium system during the iterative procedure. In practice, the potential is also always cut and shifted at a lengthscale \(r_{\text{c}}\) after each IBI iteration. For the work here, a value of \(r_{\text{c}}=8d\) was sufficient and could be lowered in some cases, e.g., when considering smaller clusters.

Implementation of IBI is accomplished through the Versatile Object-oriented Toolkit for Coarse-graining Applications (VOTCA)~\cite{doi:10.1021/ct900369w}, which is implemented with the GROMACS 4.5.3 molecular dynamics (MD) package~\cite{JCC:JCC20291}. We perform simulations comprising \(N=2048\) particles using a time step of \(dt\approx0.0003\sqrt{d^2m/(k_{B}T)}\) (\(m\) is the particle mass) and a velocity-rescale thermostat for \(T\) with characteristic time constant \(\tau=100dt\), where rescaling is done every \(10dt\). VOTCA utilizes GROMACS trajectories to calculate RDFs and potential updates accoring to Equation \ref{eqn:ibi_equation}.

\subsection{SALR model systems}

To contextualize the behaviors of the newly designed IC potentials, it is useful to make comparisons to results of an SALR interaction potential known to exhibit equilibrium cluster phases~\cite{GodfrinWagnerLiu2014,PhysRevE.91.042312,ManiBolhuis2014,ToledanoSciortino2009,Sciortino2004,LeoLue2014,Zhang2009,malins2011effect,malins2009geometric}. Specifically, we compare IC results with those from a ternary mixture model developed in a previous study~\cite{PhysRevE.91.042312} that can generate both amorphous and microcrystalline clusters. The pair potential in this model is defined as
\begin{equation} \label{eqn:salr}
\beta\varphi_{SL|i,j}(x_{i,j}) \equiv 4[\chi+(1-2\delta_{i,j})\Delta_{\chi}] ( x_{i,j}^{-2\alpha} -x_{i,j}^{-\alpha})+Q \dfrac{e^{-x_{i,j}/\xi}}{x_{i,j}/\xi}
\end{equation}
where \(x=r/d\) is a non-dimensionalized interparticle separation, \(d\) is the characteristic particle diameter, \(\chi\) quantifies a short-range attractive strength (we choose \(\alpha=100\) to set attractive wells of \(O(1\%)\) of the core diameter), and \(Q\) and \(\xi\) respectively set the magnitude and range of a long-range Yukawa repulsion. Additionally, \(\delta_{i,j}\) is the Kronecker delta, with \(i \textnormal{ (or }j)=-1,0,1\) corresponding to small, medium, and large particles, respectively; the generalized interparticle distance is defined \(x_{i,j}\equiv x-(1/2)(i+j)(\Delta_{d}/d)\). The remaining parameters are perturbative shifts to particle size \(\Delta_{d}\) and energy \(\Delta_{\chi}\), respectively. The values of \(\chi\), \(Q\), and \(\xi\) were tuned to generate SALR clusters of comparable size to the optimized ICs.

To generate amorphous cluster phases, we follow our previous work~\cite{PhysRevE.91.042312} and examine three-component mixtures that approximate suspensions with 10\% size polydispersity by using 20\% small, 60\% medium, and 20\% large particles with size perturbation \(\Delta_{d}=0.158d\), where we set \(\Delta_{\chi}=0.25\chi\) to gently promote mixing. To examine the more commonly studied monodisperse single-component model that exhibits microcrystalline clusters, we simply set \(\Delta_{d} = \Delta_{\chi}=0\).  

With the SALR model, we perform three-dimensional MD simulations of \(N=2960\) particles in the canonical ensemble with periodic boundary conditions using LAMMPS~\cite{Plimpton1995}. We use an integration time-step of \(dt=0.0005\sqrt{d^2m/(k_{B}T)}\), include interactions out to a cut-off distance of \(r_{\text{c}}=8d\), and fix temperature via a Nos\'{e}-Hoover thermostat with time-constant \(\tau=2000dt\).

\subsection{Cluster-size distribution and bond order analysis}

To characterize the instantaneous scale of the equilibrium cluster aggregates, we calculate cluster-size distributions (CSDs) quantifying the probability \(P(n)\) of observing aggregates comprising \(n\) particles. As is customary~\cite{GodfrinWagnerLiu2014,PhysRevE.91.042312,ManiBolhuis2014,ToledanoSciortino2009,Sciortino2004,LeoLue2014,Zhang2009,malins2011effect,malins2009geometric}, two particles are considered part of the same cluster if at least one of the following conditions are met: (1) their centers are within a pre-defined cutoff distance \(r_{\text{cut}}\), making them direct neighbors; and/or (2) their centers are both within \(r_{\text{cut}}\) of one (or more) other particle(s), i.e., are connected via some percolating pathway. For the IC systems, we generally use \(r_{\text{cut}} = 1.25d\) (as discussed later, results are not very sensitive to the choice of \(r_{\text{cut}}\)); for the SALR systems, we choose \(r_{\text{cut}}\) to be the range of the attractive well, which varies slightly depending on choices of parameters but is generally around \(r_{\text{cut}} \simeq 1.05d\). Throughout the remainder of the manuscript, a phase is considered clustered with aggregates of a preferred equilibrium size \(n^{*}\) based on the presence of a \emph{local maxima} in the CSD at \(n^{*}\) occurring in the range \(1 \ll n^{*} \ll N\).

Calculation of CSDs, which depend on many-body interparticle correlations, provides a practical means to access important information regarding how well clustering is reproduced in our pair-potential system. In particular, matching RDFs via IBI does not guarantee that the correct cluster size is reproduced, nor does it provide information on any undesired (i) polydispersity of emergent aggregates and (ii) free monomer or other small aggregates, both of which are absent in our constrained, target simulations.

%%%%%%%%%%%%%%%%%%%%%||%%%%%%%%%%%%%%%%%%%%%||%%%%%%%%%%%%%%%%%%%
%%%%%%%%%%%%%%%%%%%%%||ADDED PARAGRAPH ON BO||%%%%%%%%%%%%%%%%%%%
%%%%%%%%%%%%%%%%%%%%%\/%%%%%%%%%%%%%%%%%%%%%\/%%%%%%%%%%%%%%%%%%%

As a complement to the CSD analysis, we use a previously published approach~\cite{BOPcode} to calculate probability distributions \(P(x)\) where \(x=q_{4}\), \(q_{6}\), \(w_{4}\), and \(w_{6}\) are the four standard parameters characterizing local bond-orientational order (BO).~\cite{BOPdefinitions}. The comparison of these distributions between the constrained and optimized systems provides a first-order quantification of how effectively the RDF mapping preserves higher-order, local structural correlations within clusters. In calculating the local BO parameters, we employ the same \(r_{\text{cut}}\) used in the CSD calculation for identification of each particle's nearest neighbors.

%%%%%%%%%%%%%%%%%%%%%/\%%%%%%%%%%%%%%%%%%%%%/\%%%%%%%%%%%%%%%%%%%
%%%%%%%%%%%%%%%%%%%%%||ADDED PARAGRAPH ON BO||%%%%%%%%%%%%%%%%%%%
%%%%%%%%%%%%%%%%%%%%%||%%%%%%%%%%%%%%%%%%%%%||%%%%%%%%%%%%%%%%%%%

\subsection{Cluster persistence}

%*** hey Ryan: I sent you an email regarding of the rebuttal responses.

Despite the frequent measurement of static CSDs, there is little discussion in the literature regarding the dynamic stability (``lifetime'') of the contributing aggregates. 
% new
In fact, we are not aware of any generally accepted methods that quantify cluster dynamics in particle-based systems other than measurements of monomer mean-squared displacements~\cite{ClusterGlassandMSD} and dynamic structure factors~\cite{SupercooledClusterDynamics}, 
especially in ways that captures cluster persistence by explicitly incorporating dynamic bond information.
%In fact, we are not aware of any generally accepted methods by which to quantify cluster persistence in particle based systems other than measurements of mean square displacements~\cite{ClusterGlassandMSD} and dynamic structure factors~\cite{SupercooledClusterDynamics} to incorporate explicit \emph{dynamic} bond information. 
%
To facilitate such measurements, we introduce the correlation function \(\Phi(t)\), which quantifies the \emph{fractional similarity of associates in clusters} (FSAC) at an initial time \(t=0\) and a lag-time \(t>0\), where particles are \emph{associates} at a given time if they are in the same cluster. More specifically, the correlation function can be written
\begin{equation} \label{eqn:FSAC}
\Phi(t) \equiv \dfrac{1}{N} \sum_{i=1}^{N} \Bigg[ \dfrac{\Phi_{i\text{,shared}}(t)}{\Phi_{i\text{,total}}(t)}\Bigg]
\end{equation}
where \(\Phi_{i\text{,shared}}(t)\) counts the the number of particle \(i\) associates that are \emph{common} to \(t=0\) and \(t>0\) while \(\Phi_{i\text{,total}}(t)\) is the combined sum of particle \(i\) associates at \(t=0\) and \(t>0\) (without double counting particles common to both times).

For example, if particle 1 is in a cluster with particles \{2, 3, 4\} at \(t=0\), and in a cluster with particles \{2, 3, 5\} at some \(t>0\), then \(\Phi_{\text{1,total}}(t)=4\) while \(\Phi_{\text{1,shared}}(t)=2\). The fractional similarity of associates to particle \(1\) is then \(\Phi_{\text{1,shared}}(t)/\Phi_{\text{1,total}}(t)=0.5\). In the special case that particle \(i\) is a monomer at both time-points we assume \(\Phi_{i\text{,shared}}(t)/\Phi_{i\text{,total}}(t)=1\), thus ensuring monomers that remain as monomers contribute positively to the score. Averaging the score of every particle then yields Equation \ref{eqn:FSAC}.

The correlation function has the range \(\Phi(t) = [0,1]\), where \(\Phi(t>0) = 1\) means that all clusters (including monomers) contain the same particles at both time-points and \(\Phi(t) = 0\) means that all particles possess temporally exclusive sets of associates. In a macroscopic system with finite-sized clusters, \(\Phi(t\rightarrow\infty)=0\) due to single-particle and cluster diffusion. However, this is not captured in a finite-sized box as particles can wrap through the periodic boundaries. As such, we include an additional rule: if two particles \(i\) and \(j\) ever move further than a half box-length away from one another at \(t>0\), each is subsequently treated as a new, previously unrecognized particle with respect to the other at all future time-points \(t'>t\). Thus, two associates at \(t=0\) that diffuse \emph{very far} from one another before becoming associates again (in a potentially arbitrary cluster) are not counted as temporally common. 

Altogether, the counts in Equation \ref{eqn:FSAC} can be expressed more formally as \(\Phi_{i\text{,shared}}(t)=\sum_{j\neq i} \Theta_{ij}(0)\Theta_{ij}(t)\gamma_{ij}(t)\) and \(\Phi_{i\text{,total}}(t)=\sum_{j\neq i} [\Theta_{ij}(0) + \Theta_{ij}(t)] - \Phi_{i\text{,shared}}(t)\) where

\begin{equation*}
    \Theta_{ij}(t)\equiv
    \begin{cases}
      1, & \text{\(i\) and \(j\) in same cluster at lag-time \(t\)} \\
      0, & \text{otherwise}
    \end{cases}
\end{equation*}

\noindent where the instantaneous cluster analysis at each time is done in the usual way. The factor \(\gamma_{ij}(t)\) enforces the rule concerning temporal pairwise drift and is given by

\begin{equation*}
    \gamma_{ij}(t)\equiv
    \begin{cases}
      1, & \forall\text{ }t' \leq t, r_{ij}(t') \leq L/2 \\
      0, & \text{otherwise}
    \end{cases}
\end{equation*}

\noindent where \(L\) is the simulation box length and \(r_{ij}(t) \equiv \|\textbf{r}^{\text{w}}_{ij}(0) + \Delta \textbf{r}^{\text{u}}_{ij}(t)\|\) with \(\textbf{r}^{\text{w}}_{ij}(0)\) the \emph{wrapped} initial displacement vector between particles \(i\) and \(j\) and \(\Delta \textbf{r}^{\text{u}}_{ij}(t)\) the corresponding net cumulative \emph{unwrapped} displacement over lag-time \(t\).

%%%%%%%%%%%%%%%%%%%%%%%%%%%%%%%%%%%%%%%%%%%%%%%%%%%%%%%%%%%%%%%%%%%%%%%%%%%%%%%%%%%%%%%%%%%%%%%%%%%%%%%

\section{Results and discussion}

\subsection{Cluster structure}

In Fig.~\ref{sch:FigureRDFs}, we compare RDFs obtained from the constrained Monte Carlo simulations to those of MD simulations using the IBI-optimized potentials. The \(g(r)\) profiles show very good agreement over three orders of magnitude (demonstrating the successful application of the IBI approach), and also--as expected--exhibit features that are consistent with clustering and atypical of simple fluids. One such feature is the highly structured, liquid droplet envelope extending over multiple particle diameters. This droplet region is terminated by a particle-rarefied window, which, in a highly averaged sense, defines the cluster center-to-surface distance and provides an intuitive division of \(g_{\text{tgt}}(r)\) into \emph{intracluster} and \emph{intercluster} particle correlations. Intercluster \(g_{\text{tgt}}(r)\) correlations are oscillatory, with relatively long characteristic wavelengths set by the effective cluster size \(d_{\text{eff}}\), which is originally encoded during the reference constained MC simulations (see Section 2.1). Importantly, \(d_{\text{eff}}\) also controls the depth of the depletion window in \(g_{\text{tgt}}(r)\), where the depletion depth is greater when the repulsive-shell lengthscale (\(d_{\text{eff}}\)) is larger. Intercluster repulsion at the COM level is essential towards achieving convergence in the IBI scheme, as in its absence, we find that the intermediate IBI steps become unstable towards large scale aggregation. In addition, the thickness of the repulsive layer needed for convergence grows with cluster size (see Section 3.4).

\begin{figure*}[ht!]
\begin{center}
  \includegraphics{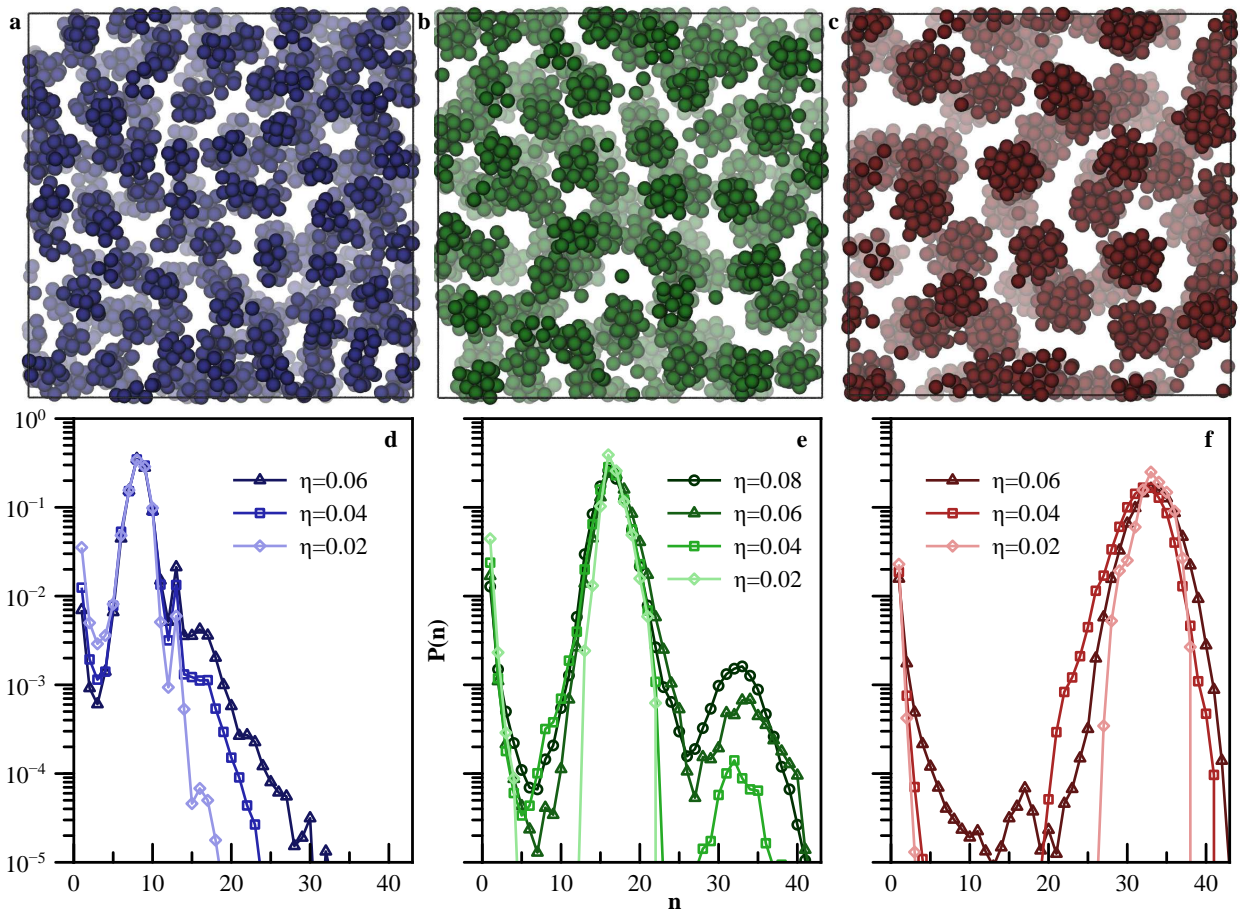}
  \caption{(a-c) Simulation snapshots for \(n_{\text{tgt}}=8\), 16, and 32, respectively, at a volume fraction of \(\eta=0.06\). (d-f) Cluster size distributions for \(n_{\text{tgt}}=8\), 16, and 32, respectively, for various \(\eta\).}
  \label{sch:FigureSNAPSHOTS}
\end{center}
\end{figure*}

\begin{table}[ht]
\centering
\begin{tabular}{  >{\centering\arraybackslash}m{1.2cm} |  >{\centering\arraybackslash}m{1.2cm}  >{\centering\arraybackslash}m{1.2cm}  >{\centering\arraybackslash}m{1.2cm}  >{\centering\arraybackslash}m{1.2cm} } 
\hline
\(\eta\) & 0.02 & 0.04 & 0.06 & 0.08\\
\hline
\(n^{*}\) & 8 & 8 & 8 & \\ 
\(\langle n \rangle\) & 7.95 & 8.26 & 8.48 & \\ 
\(\delta n\) & 2 & 2 & 2 & \\
\hline
\(n^{*}\) & 16 & 16 & 16 & 16 \\
\(\langle n \rangle\) & 15.90 & 16.03 & 16.65 & 16.25\\ 
\(\delta n\) & 3 & 3 & 3 & 3 \\
\hline
\(n^{*}\) & 33 & 32 & 33 & \\ 
\(\langle n \rangle\) & 32.68 & 31.46 & 32.66 & \\ 
\(\delta n\) & 4 & 4 & 5 & \\
\hline
\end{tabular}
\caption{Average cluster size and polydispersity measures for all cluster sizes (\(n_{\text{tgt}}=8\), 16 and 32) and volume fractions.}
\end{table}

Of course, convergence in \(g(r)\) alone does not guarantee preservation of higher-order particle correlations, which \emph{could}~\footnote[3]{Matching higher-order correlations is not strictly necessary for strong clustering so long as the resultant set of many body correlations generated by the optimized pair potential correspond to strongly clustered states.} %Explicit verification of such a scenario would require complicated many body correlation calculations beyond the scope of the work here aimed only at discerning if strong clustering can be preserved (ideally suited to a CSD analysis).} % !!!%%% JAB removed this portion as I feel the bond-orientational order analysis does being to address this, and thus this is within `the scope' of the study.
play an important role in clustering behavior. 
%revised JAB just to be more succinct
However, we find that strong clustering emerges using the designed pairwise potentials, as evidenced by the representative simulation snapshots shown in Figs.~\ref{sch:FigureSNAPSHOTS}(a-c), the three corresponding CSDs in Figs.~\ref{sch:FigureSNAPSHOTS}(d-f), and simulation movies of these systems (provided in the Supplementary Material). 
% old
%However, Fig.~\ref{sch:FigureSNAPSHOTS} and the supplemental movies demonstrate that strong clustering still emerges using the designed pairwise potentials [see supplementary material for potentials], as is evident from the representative simulation snapshots shown in Figs.~\ref{sch:FigureSNAPSHOTS}(a-c) [each has a supplementary movie], and the three corresponding CSDs, in Figs.~\ref{sch:FigureSNAPSHOTS}(d-f). 
Visual inspection of Figs.~\ref{sch:FigureSNAPSHOTS}(a-c) reveals that the self-assembled clusters are (1) highly-distinguishable (well-defined); (2) spherical and droplet-like; and (3) similar in size to the enforced analogs in Fig.~\ref{sch:FigureRDFs}. The CSDs, calculated using \(r_{\text{cut}}=1.25d\)\footnote[1]{We note that the appropriate choice of \(r_{\text{cut}}\) is non-unique given the shapes of \(u(r)\) (see Section 3.4); however, the clusters are sufficiently well-defined such that various cutoffs \(r_{\text{cut}}/d=[1.1,1.5]\) yield negligible differences in the CSDs.}, % end footnote
confirm that the optimized potentials promote clusters of the desired sizes, as quantified by both the characteristic maximum \(n^{*}\)
\begin{equation} \label{model:1}
P(n^{*}) \equiv \textnormal{max}_{n}P(n)
\end{equation}
and the average value
\begin{equation} \label{model:1}
\langle n \rangle \equiv \sum_{n=1}^{\infty} nP(n)
\end{equation}
both listed in Table 1. We also note that the clusters are so well defined that peaks corresponding to infrequently connected clusters (two times primary cluster size) are also observed at higher \(\eta\) (not visible for \(n_{\text{tgt}}=32\)).

To complement the size measures \(n^{*}\) and \(\langle n \rangle\), we also calculate the peak-width \(\delta n\) according to
\begin{equation} \label{model:1}
0.90 \leq \sum_{n=n_{\text{tgt}}-\delta n}^{n_{\text{tgt}}+\delta n} P(n)
\end{equation}
which is the (one-sided) range in \(n\) about \(n^{*}\) that accounts for 90\% of the summated \(P(n)\) curve. As demonstrated in Table 1, \(\delta n\) is of a reasonable size for all \(n_{\text{tgt}}\), with the ratio \(\delta n/n^{*}\) decreasing with increasing \(n^{*}\) (intuitively, \(\delta n/n^{*} \rightarrow 0\) in the thermodynamic, cluster-size limit). Finally, as is clear from Figs.~\ref{sch:FigureSNAPSHOTS}(d-f), these highly monodisperse clusters also coexist with a small but numerically detectable fraction of free monomer, which can also be visually gleaned from Figs.~\ref{sch:FigureSNAPSHOTS}(a-c) and the three corresponding supplemental movies. This bifurcation of the system into two primary populations suggests a similarity to liquid-vapor coexistence, albeit on the microscale.

\begin{figure}
  \includegraphics{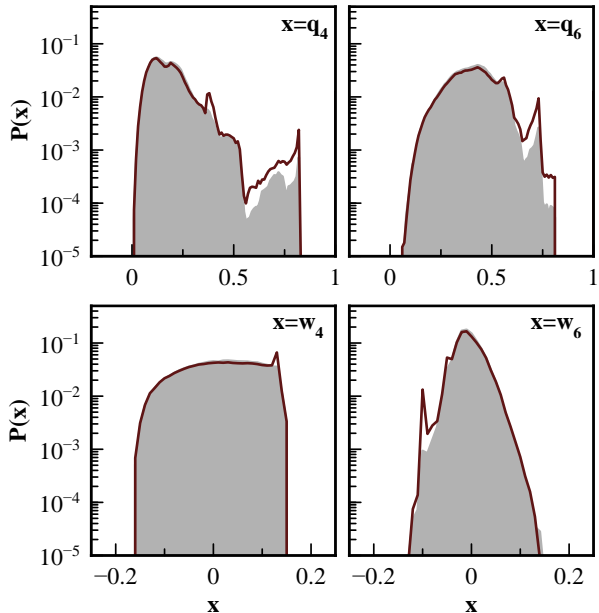}
  \caption{Bond-orientational order probability distributions for \(n_{\text{tgt}}=32\) at packing fraction \(\eta=0.06\). The filled grey and open red curves indicate constrained and optimized simulations (using clusters with sizes between \(n^{*}\pm \delta n\)), respectively. Analogous calculations for the two other cluster sizes are available in the Supplementary Material.}
  \label{sch:FigureBO}
\end{figure}

%%%%%%%%%%%%%%%%%%%%%||%%%%%%%%%%%%%%%%%%%%%||%%%%%%%%%%%%%%%%%%%
%%%%%%%%%%%%%%%%%%%%%||ADDED PARAGRAPH ON BO||%%%%%%%%%%%%%%%%%%%
%%%%%%%%%%%%%%%%%%%%%\/%%%%%%%%%%%%%%%%%%%%%\/%%%%%%%%%%%%%%%%%%%

While the CSD calculations clearly demonstrate that target cluster sizes are reproduced and that minimal intrusive smaller or larger objects are present, we also find that reproducing pair structure carries over to some higher-order structural correlations. Here, we specifically consider the local higher-order bond-orientational order (BO) parameters~\cite{BOPdefinitions,BOPcode} \(q_{4}\), \(q_{6}\), \(w_{4}\), and \(w_{6}\), which are understood to be strongly correlated with pair structure for \emph{homogeneous} disordered liquids and even glassy packings~\cite{OrderInHardSpherePackings,OrderInLJFluid}. 
%
%Thus, we expect intracluster BO of the optimized systems to be well-mapped by matching RDFs--something that is conveniently tested via probability distributions, \(P(x)\), of the four popular BO parameters, \(x=q_{4}\), \(q_{6}\), \(w_{4}\) and \(w_{6}\). 
%
In Fig.~\ref{sch:FigureBO}, we compare probability distributions \(P(x)\) of these local BO parameters for constrained and optimized systems of \(n_{\text{tgt}}=32\) clusters at \(\eta=0.06\). For the latter case, a distribution of cluster sizes is present. Therefore, to compare clusters of similar size, we collect statistics over clusters of instantaneous size \(n^{*}\pm \delta n\) (see Table 1) from the optimized simulations. The intracluster higher-order correlations are indeed well preserved between two cases, though we note that the observed agreement slightly diminishes as target cluster size decreases (see Supplementary Material for \(n_{\text{tgt}}=8\) and \(16\) calculations), a trend likely related to the issues we discuss in the following paragraph.
%
%As shown in Fig.~\ref{sch:FigureBO}, local intracluster correlations present in the [\(\eta=0.06\), \(n_{\text{tgt}}=32\)] constrained system are very well preserved in the optimized analog where only clusters with a size falling between \(n^{*}\pm \delta n\) (see Table 1) were included. This observed agreement systematically diminishes, though not greatly, as the target cluster size gets smaller (see SI for analogous \(n_{\text{tgt}}=8\) and \(16\) calculations). The growing discrepancies between the BO of the constrained and optimized systems for smaller clusters may be related to the issues discussed in the following paragraph.

%%%%%%%%%%%%%%%%%%%%%/\%%%%%%%%%%%%%%%%%%%%%/\%%%%%%%%%%%%%%%%%%%
%%%%%%%%%%%%%%%%%%%%%||ADDED PARAGRAPH ON BO||%%%%%%%%%%%%%%%%%%%
%%%%%%%%%%%%%%%%%%%%%||%%%%%%%%%%%%%%%%%%%%%||%%%%%%%%%%%%%%%%%%%

Beyond the cluster sizes (\(8 \leq n_{\text{tgt}} \leq 32\)) considered in Figs.~\ref{sch:FigureRDFs}-~\ref{sch:FigureSNAPSHOTS}, we also attempted the IBI approach to obtain pair potentials \(u(r)\) that would generate smaller and larger amorphous clusters; however, various challenges emerge in either limit. For clusters of size \(n_{\text{tgt}} < 8\), we find that it is difficult for an isotropic pair potential to generate well-differentiated amorphous clusters while simultaneously suppressing similarly sized, \emph{ordered}, intracluster configurations that cannot be easily penalized on the basis of single length-scale. This type of issue arises in a minor way even for \(n_{\text{tgt}} = 8\), as evidenced by the small \(n=13\) peak in the CSDs of Fig.~\ref{sch:FigureSNAPSHOTS}(d), which corresponds to clusters with a central seed and 12 closely-packed neighbors (the sphere kissing number~\cite{conway2013sphere}). We found that the intrusion of ordered off-target clusters is most prevalent when attempting to stabilize dimers (\(n_{\text{tgt}}=2\)), where 3-mers and 4-mers were always more highly favored. This is a result of the minimalistic fluctuation (single particle) being of order the size of the desired cluster (dimer) and the fact that 3-mers and 4-mers can assemble into objects with no discriminatory length scales (triangles and pyramids respectively) for a pair potential to disfavor. More generally though, \emph{larger} (than the target size) intrusive clusters must be ordered and closely packed so as to utilize space more efficiently and avoid sampling the growth limiting repulsions present in the potentials (discussed in Section 3.4) optimized for the less space efficient amorphous clusters.

On the other hand, attempts to generate potentials that stabilize fluids of \(n_{\text{tgt}}=64\) failed to converge for all \(\eta \geq 0.02\). It is unclear whether ICs could be designed for clusters of this size with a different choice of parameters in the target simulation (e.g., we found that increasing \(d_{\text{eff}}\) resulted in a somewhat more stable, though ultimately unsuccessful, optimization) or if this failure is symptomatic of a fundamental limitation of a pair potential to stabilize fluids of large ICs. Exploring and articulating the limits of a pair potential to create given fluid architectures remains an open question for future research.

\subsection{Dependence on density} 

\begin{figure}
  \includegraphics{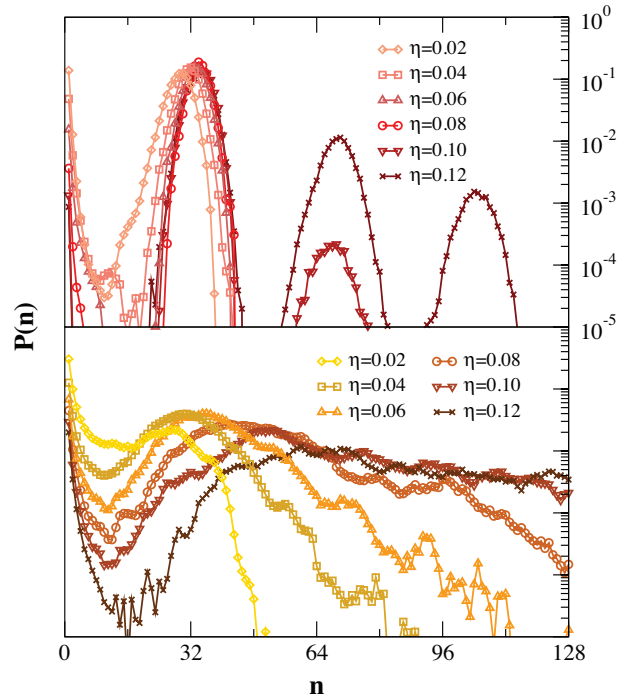}
  \caption{(\emph{upper}) CSDs corresponding the potential optimized for \(n_{\text{tgt}}=32\) at \(\eta=0.06\) at volume fractions ranging from \(0.02\rightarrow0.12\). (\emph{lower}) CSDs calculated for an SALR potential (\(\chi=5.7\), \(Q=0.2\), \(\xi=2.0\)) that yields clusters with size \(n^{*} \approx 32\) at \(\eta=0.06\), used for the same series of volume fractions.}
  \label{sch:FigureCOMPRESSION}
\end{figure}

While the results above demonstrate that the IBI optimization generates pair potentials that induce the desired clustering, we next demonstrate that these potentials are also robust with respect to changes in \(\eta\) via two complementary approaches. The first is a comparison of the potentials optimized at specific values of \(\eta\) in the supplementary material. Overall insensitivity, including the functional form, to density is found across all cluster sizes--only a weak decrease in the overall amplitude of the potentials with increasing \(\eta\) is observed. The second demonstration of insensitivity to density is demonstrated by simulating optimized potentials generated for a particular \(\eta\) under either a slow (quasi-equilibrium) expansion or compression. In the top panel of Fig.~\ref{sch:FigureCOMPRESSION}, we plot CSDs for simulations of the \(u(r)\) potential corresponding to \(n_{\text{tgt}}=32\) at \(\eta=0.06\) at various terminal (equilibrated via long runs between compressions) packing fractions \(\eta=[0.02,0.12]\), where it is apparent that the CSDs possess primary peaks (\(30<n^{*}<35\)) centered near the original targeted value.

Consistent with the notion that free monomer particles represent the ``vapor'' in a microscopic liquid-vapor coexistence, as \(\eta\) decreases, the integrated amount of monomer and other small aggregates increases (as expected from energy-entropy compensation arguments) and the main peak in the CSD shifts left relative to the original position.  Notably, at the largest packing fractions, \(\eta=0.10\) and 0.12, secondary peaks emerge with local maxima at \(n \approx 2n^{*}\), which can be attributed to configurations where at least two particles from neighboring clusters come within \(r_{\text{cut}}\) of one another with some frequency. In fact, for \(\eta=0.12\), there are multiple peaks in the CSD at appoximate intervals of \(n^{*}\) extending out to \(n \approx 6n^{*}\) clusters, where peak height is negatively correlated with \(n\).

The robustness of \(n^{*}\) and the appearance of features in the CSDs at intervals of \(n^{*}\) upon compression indicate that the clusters in the IC systems remain well differentiated and of the preferred size despite greater proximity (and close contact). This is in contrast to analogous CSD measurements for SALR mixtures that form fluid (non-crystalline) clusters of similar characteristic size at \(\eta=0.06\), which are shown in the bottom panel of Fig.~\ref{sch:FigureCOMPRESSION}. (The SALR parameters were chosen to produce comparably sized clusters, \(n^{*}=35\), at \(\eta=0.06\)). First, it is clear that, at the reference volume fraction \(\eta=0.06\), the SALR fluid has a considerably broader distribution of cluster sizes than the IC fluid, and monomer remains the dominant species (this is true for any given reference \(\eta\), except when generating highly arrested percolating gel states). Moreover, the \(n^{*}\) peak is considerably more sensitive to changing \(\eta\) than in the IC systems, and for \(\eta > 0.06\), there is an increasingly wide distribution of competitive cluster sizes. In other words, the clusters undergo continuous and unorganized growth with poor cluster distinguishability. 

These dichotomies in cluster size-specificity and differentiation at higher \(\eta\) between IC and SALR systems exhibiting amorphous clusters coincides with another behavioral difference: the clusters in the IC systems, while remaining as distinguishable entities with intracluster fluidity, self-organize at the COM level into crystalline superlattices for \(\eta \geq 0.08\), indicating the density range where the liquid state of clusters becomes thermodynamically unfavorable (or even unstable).\footnote[2]{Based on preliminary investigation, the crystalline superlattice types we observe are sensitive to system size, as is expected given the small number of clusters in most of the systems we examine (e.g., \(N_{\text{cluster}} = 64\) for \(N=2048\) and \(n_{\text{tgt}} = 32\)). Further research on the crystal phase, while beyond the scope of this initial fluid-state study, is an interesting topic for future work.} 
In contrast, the SALR fluid remains disordered at the cluster level for all \(\eta\). The ability of the amorphous clusters in the IC system to self-assemble into a lattice strongly supports the interpretation of clusters as renormalized entities. Superlattice formation also attests to the notable monodispersity posessed by the clusters as the presence of polydispersity inhibits crystalline phases (for kinetic and/or thermodynamic reasons~\cite{PhysRevLett.103.135704,pusey2009hard}).

While the \(\eta\) at which the IC clusters form superlattices is, at first glance, quite low, we note that the effective packing fractions \(\eta_{\text{eff}}\) of the whole clusters (treating them as renormalized objects; see Section 2.1) are considerably higher. For monodisperse clusters of \(n^{*}=32\), \(\eta_{\text{eff}}\approx0.44\) at \(\eta=0.08\) and \(\eta_{\text{eff}}\approx0.55\) at \(\eta=0.10\), conditions which approximately correspond to those at which crystallization is induced in simple fluids dominated by steep interparticle repulsions~\cite{PhysRevLett.103.135704,pusey2009hard}. Incidentally, this tendency toward COM crystallization makes it difficult to obtain convergence in the IBI scheme at similarly high effective packing fractions.

\subsection{Cluster persistence and particle motions}

\begin{figure}
  \includegraphics[trim={0 3.2cm 0 0},clip]{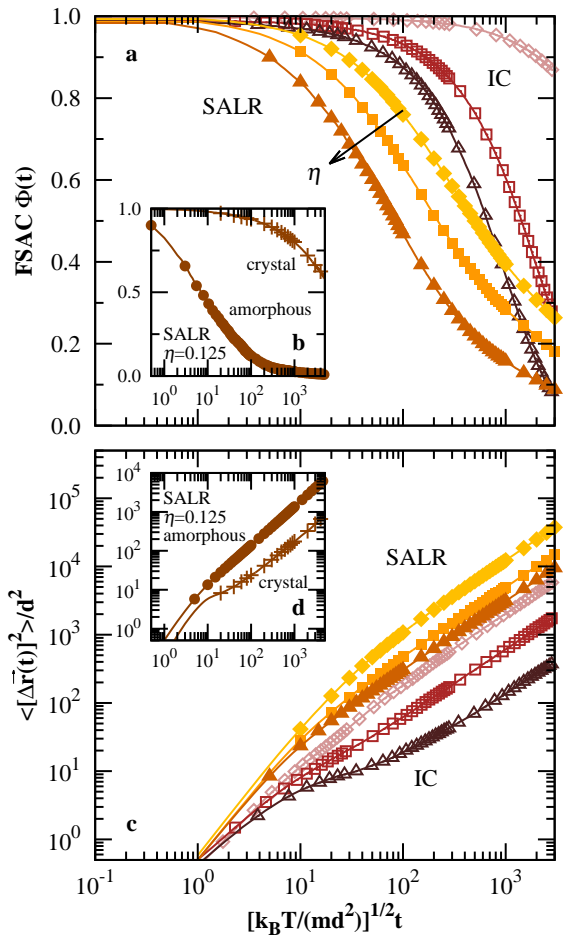}
  \caption{(a) Cluster persistence (FSAC) correlation function \(\Phi(t)\) for \(n^{*}=32\) IC systems (unfilled symbols) and \(n^{*}\approx32\) SALR mixtures (filled symbols) at packing fractions \(\eta\) = 0.02 (diamonds), 0.04 (squares), and 0.06 (triangles). The attractive strengths \(\chi\) in the Equation \ref{eqn:salr} potentials at the three packing fractions are \(\chi\) = 6.1, 5.9, and 5.7, respectively, and the repulsions are defined by \(Q=0.2\) and \(\xi = 2.0\). (c) Single-particle mean-squared displacements (MSDs) for the same IC and SALR systems in (a). Insets (b) and (d) compare data for amorphous (mixture) and microcrystallizing (single-component) SALR models with \(n^{*} \approx 32\), where the potentials are defined by \(\chi = 5.4\), \(Q = 0.2\), \(\xi = 2.0\) and \(\chi = 6.0\), \(Q = 0.5\), \(\xi = 2.0\), respectively. Note that for visual clarity, lines trace all available data in each panel while symbols do not.}
  \label{sch:FigureFSAC}
\end{figure}

In Fig.~\ref{sch:FigureFSAC}, we consider the temporal cluster persistence and single-particle (i.e., monomer) dynamics of IC (\(n^{*}=32\) and SALR \(n^{*} \approx 32\) systems, where the FSAC correlation function profiles \(\Phi(t)\) quantitatively demonstrate that the IC systems exhibit signficantly longer cluster ``lifetimes'' than their SALR counterparts. To wit, as shown in Fig.~\ref{sch:FigureFSAC}(a), the half-life values \(t_{1/2}\) (i.e., times at which \(\Phi(t) = 0.5\)) of the IC systems are approximately an order of magnitude or more greater than the values for SALR systems for given packing fractions. We consider this strong cluster fidelity a byproduct of generating highly monodisperse spherical clusters (and vice versa) like those illustrated in Fig.~\ref{sch:FigureSNAPSHOTS}. If, instead, clusters are frequently exchanging constituents with each other or the monomer ``vapor'', they will tend to exhibit rather highly fluctuating (and/or instantaneously non-spherical) interfaces. Thus, instantaneous configurations will appear less superficially monodisperse and CSD profiles will be necessarily broader. This notion is consistent with the CSDs in Fig.~\ref{sch:FigureCOMPRESSION}, where the less exchange-prone IC systems exhibit much greater size-specificity than the SALR mixtures (one can also compare the cluster snapshots in Fig.~\ref{sch:FigureSNAPSHOTS} with those in Fig. 4 of a previous publication~\cite{PhysRevE.91.042312}). 

In terms of density dependence, we observe that cluster persistence is negatively correlated with \(\eta\) for all \(n_{\text{tgt}}\) cluster sizes considered, as typified by the \(n_{\text{tgt}}=32\) results shown in Fig.~\ref{sch:FigureFSAC}. This is easily understood by considering that increasing \(\eta\) necessarily places clusters in close proximity, where their internal density fluctuations help facilitate the transfer of monomers between clusters. This qualitative effect is, of course, relevant for both IC and SALR systems, though the precise nature of the density-dependence differs due to the relative monodispersity and sphericity of the IC clusters.

However, an additional consideration when comparing the persistence and monodisperity of IC and SALR clusters is whether the SALR systems are allowed to form microcrystalline clusters (a phase change), as opposed to the amorphous clusters considered in Fig.~\ref{sch:FigureFSAC}(a).
In prior work, we showed via time-lag snapshots that single-component SALR models tend to form crystalline clusters with greater temporal persistence than slightly size-polydisperse (at the monomer level) mixtures that thwart crystallization~\cite{PhysRevE.91.042312}. In Fig.~\ref{sch:FigureFSAC}(b), this is shown quantitatively for \(n^{*} \approx 32\) clusters at \(\eta = 0.125\), where we find the single-component SALR clusters exhibit \(t_{1/2}\) values orders of magnitude longer than amorphous clusters of the SALR mixtures. These crystallized clusters also display half-lives much longer than even their IC counterparts: for example, recalling that increasing density accelerates the FSAC decay rate, the single-component \(n^{*} \approx 32\) SALR clusters at \(\eta = 0.125\) exhibit \(t_{1/2} \approx 10^{4}\), which is comparable to that of \(n^{*}=32\) IC clusters at only \(\eta = 0.04\). %RBJ: this statement is not needed I feel. Thus, I used another version that talks about microcrystallinity (touching on Delias comments below) %However, despite the fact that the crystalline clusters can persist for very long times, the more straightforward comparison between amorphous IC and SALR systems nonetheless indicates that the IC systems show much greater cluster monodispersity and persistence given fluid-like intracluster rearrangements 
Given the orders of magnitude discrepancy in \(t_{1/2}\) between the single-component and mixture SALR models, it seems reasonable to ascribe the qualitative differences in cluster monodispersity and persistence to the microcrystallinity in the former, as opposed to the modest perturbation to the liquid state resulting from the weak polydispersity that we have employed.~\footnote[3]{Note that slight polydispersity of the constituent monomers does not preclude the formation of highly monodisperse clusters, as evidenced by a many-body model that generates ICs~\cite{Nguyen23062015}.}

Comparing Figs.~\ref{sch:FigureFSAC}(a) and \ref{sch:FigureFSAC}(c), we observe that trends in amorphous cluster persistence are complemented by the single-particle mean-squared displacement (MSD) profiles, where particle motions comprise both intracluster diffusion and slaved motion due to diffusion of entire clusters. First, we note that the IC systems show slower single-particle dynamics relative to SALR mixtures, presumably because the slaved-motion effect persists out to longer timescales. Interestingly, for the IC systems, we also find the emergence of transient plateaus in the MSD profiles by \(\eta = 0.06\) for \(n_{\text{tgt}}=32\) (and slightly higher \(\eta\) for smaller \(n_{\text{tgt}}\)). (A similar plateau also emerges for the highly persistent clusters associated with the single-component microcrystallizing SALR system; see Fig.~\ref{sch:FigureFSAC}(d).) Such a feature is typically observed in the context of cooperative glassy single-particle dynamics in dense and or supercooled fluids~\cite{PhysRevLett.103.135704,pusey2009hard}. For these systems, on the other hand, it is intuitive that this signature is a consequence of the whole-cluster \(\eta_{\text{eff}}\) being much larger than \(\eta\) for systems of well-defined spherical clusters (as described in Section 3.2). 

Thus, even for rather low packing fractions \(\eta < 0.08\), the presence of persistent and highly distinguishable clusters of an appreciable size (e.g., \(n^{*}>16\)) drives the emergence of cooperative whole-cluster dynamics, which is then reflected on the single-particle level. In other words, the \(\eta\)-range over which ICs truly diffuse around one another in a fluid-like manner \emph{is quite small}. In contrast, no such shoulders in the MSDs emerge for the amorphous SALR cluster phases up to and above \(\eta = 0.20\) (not shown), pointing to the quite rapid exchange and reformulation of clusters in non-crystallizing SALR fluids (provided the short-range attractions are not so strong as to generate dynamically arrested gel phases).

\subsection{Optimized potentials}

\begin{figure}
  \includegraphics{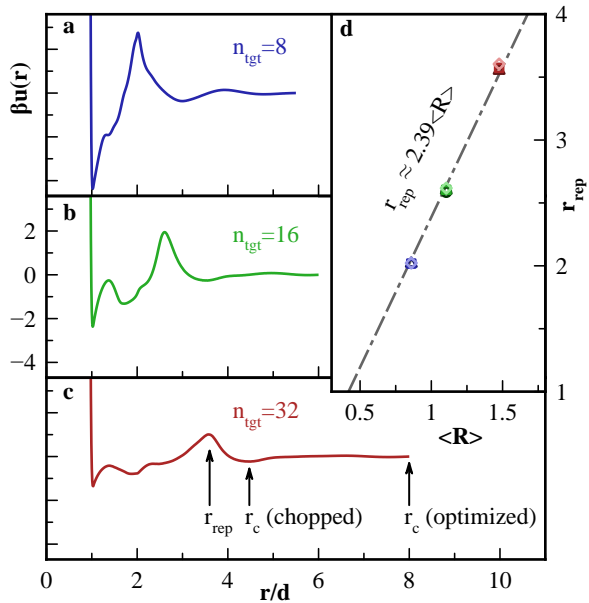}
  \caption{(a-c) Representative examples at \(\eta=0.04\) of the IBI optimized pair potentials which yield the target radial distribution functions for \(n=\)8, 16 and 32 respectively. (d) Optimized potential \(r_{\text{rep}}\) values and their corresponding target simulation average radii of gyration \(\langle R \rangle\) values for all densities studied. The linear fit (as indicated by the dashed line) is constrained such that \(r_{\text{rep}}(0)=0\).}
  \label{sch:FigurePOTENTIALS}
\end{figure}

In Fig.~\ref{sch:FigurePOTENTIALS}, we turn our attention to the pair potentials \(u(r)\) that result from the IBI optimization to yield ICs, showing particular examples for various \(n_{\text{tgt}}\) at \(\eta=0.04\) (all 10 potentials--3 for \(n_{\text{tgt}}=\)8 and 32, and 4 for \(n_{\text{tgt}}=16\)--are provided in Supplementary Material). The sensitivity of the potentials to density is minimal\footnote[4]{Increasing density leads to a small, overall amplitude decrease in the pair-potential. This insensitivity is nontrival when the effective volume fraction, \(\eta_{\text{eff}}\), for each of the target clusters is considered. A true \(\eta\) range of \(0.02\rightarrow0.6\) corresponds to \(0.08\rightarrow0.24\) for \(n=8\) and \(16\) and \(0.11\rightarrow0.33\) for \(n=32\).} and, in all cases, the potential is dominated by a broad attractive basin terminated by a repulsive barrier that falls off quickly about its maximum at \(r_{\text{rep}}\). As demonstrated in Fig.~\ref{sch:FigurePOTENTIALS}(d), the lengthscale \(r_{\text{rep}}\) is intimately related to the cluster size, as it is directly proportional to the average cluster radius of gyration \(\langle R \rangle\) extracted from the constrained simulations and has virtually no dependence on \(\eta\) for a fixed cluster size. Interestingly, this direct proprotionality between \(r_{\text{rep}}\) and \(\langle R \rangle\) incorporates the physically intuitive constraint that \(r_{\text{rep}}\) must vanish when \(\langle R \rangle = 0\). This implies that the cluster sizes \(8 \leq n^{*} \leq 32\) fall in the ``large-cluster'' asymptotic limit where the discrete nature of the particles comprising the clusters is unimportant.

\begin{figure}
  \includegraphics{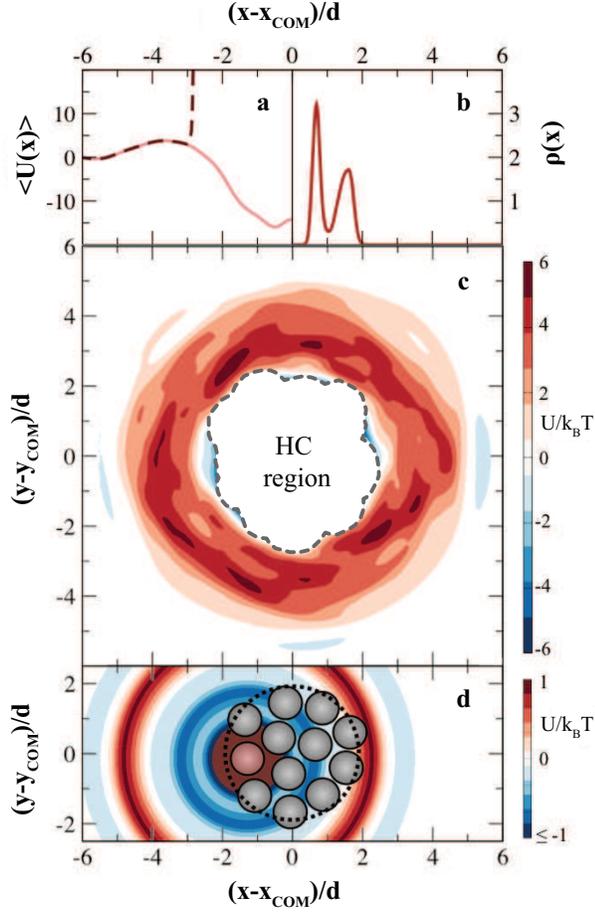}
  \caption{(a) Average potential energy for interaction of a cluster \(n=32\) with a single particle with (dashed) and without (solid) the HC component of the IBI-optimized potential. (b) Radial density profile for a single cluster. (c) A two-dimensional slice of the potential energy corresponding to a representative cluster. (d) Schematic of a cluster with the optimized potential (\(n=32\)) overlaid for a selected particle. The outer rim of the cluster (dotted circle) has a radius of 2d based on the intercluster radial distribution function (panel b). Data in panels (a-c) are generated by a simulation of a single cluster (\(n=32\)) with a constrained radius of gyration, and the data in panels (a-b) are radially averaged.}
  \label{sch:FigureCLUSTERPOTENERGY}
\end{figure}

To understand how these pair potentials strongly limit cluster growth to create renormalized objects, we turn to Fig.~\ref{sch:FigureCLUSTERPOTENERGY}, where we show that the broad attractive wells drive local densification while the repulsive barriers collectively generate strong repulsive coronas around the aggregates. Fig.~\ref{sch:FigureCLUSTERPOTENERGY}(a) shows the average potential energy that a particle placed in or near an \(n=32\) cluster experiences as a function of distance from the cluster COM using the pair potential optimized at \(\eta= 0.06\). We calculate this interaction by averaging over particle positions from a simulation of an isolated cluster and either include or exclude the hard-core component (\(r < d\)) from the calculation, where the latter is done to provide a \emph{highly averaged} measurement for the intracluster environment. From the former, it is clear that before any particle approaches the cluster sufficiently closely to sample the very steep effective potential derived from excluded volume effects, it ``sees'' an additional repulsive barrier at larger \(r\) that, in effect, terminates growth. From the latter, we see that the particles which comprise the cluster are situated within a spherical attractive region by way of comparision to the extent of the radial density profile of the cluster (Fig.~\ref{sch:FigureCLUSTERPOTENERGY}(b)). 

In Fig.~\ref{sch:FigureCLUSTERPOTENERGY}(c), we show a representative two-dimensional slice of the potential energy landscape for a \emph{single} \(n=32\) cluster configuration, which illustrates that the repulsive corona is instantaneously quite strong (\(2k_{\text{B}}T \leq U(x,y) \leq 6k_{\text{B}}T\)) and has a thickness comparable to the cluster radius. This latter observation can be understood as follows: Fig.~\ref{sch:FigureCLUSTERPOTENERGY}(d) shows an appropriately scaled schematic of a cluster that is also aligned with the heat map in panel (c), where we show the potential due to the red-colored particle as a heat map. To the right, the repulsive barrier of the highlighted particle roughly coincides with the opposite edge of the cluster, building in size specificity. However, the isotropic nature of the pair potential necessitates that this same particle also contributes repulsions in the opposite direction, where its repulsive barrier approximately coincides with the ``outer'' edge of the repulsive corona. This slaving of the outermost cluster repulsion range to cluster size may explain an outcome of the IBI optimizations: as \(n_{\text{tgt}}\) was increased, thicker protective shells surrounding the clusters (as quantified by \(d_{\text{eff}}\)) were required in the constrained simulations in order to stabilize the subsequent IBI scheme. While not precluding the existence of cluster forming pair interactions that do not obey such size-thickness slaving; the above interpretation suggestive that the most ``reasonable'' pair interactions capable of generating clusters do.

Interestingly, our pair potentials may be experimentally realizable via charged-monolayer protected gold nanoparticles, as demonstrated by Alexander-Katz and coworkers~\cite{Katz}. At the level of the pair free energy change between two such particles along a radial coordinate, they found from van der Waals, electrostatic, phobic and entropy contributions that a wide attractive basin followed by a repulsive hump can be realized. Importantly many tunable parameters exist in this experimentally realizable system making this a promising avenue for future research.

\subsection{Influence of long-range interactions}

While the main features of the potentials occur on the order of a few particle diameters in length, all of the optimized pair potentials also possess weak, oscillating, longer ranged tails. In the interest of simplifying the optimized potentials, we consider the impact of eliminating these tails by cutting and shifting the \(n_{\text{tgt}}=16\) optimized potentials at the first minimum beyond the main repulsive hump (as labeled in Fig.~\ref{sch:FigurePOTENTIALS}(c)).\footnote[5]{One cannot simply arrive at a chopped potential by applying IBI with an arbitrarily short cutoff. Any RDF that results from optimizing with a poorly chosen cutoff need not even satisfactorily match the target RDF because the method will be ``ill-conditioned''. On the other hand, truncating the fully optimized potential acts like a perturbation to the ``correct'' result.} 

\begin{figure}
  \includegraphics{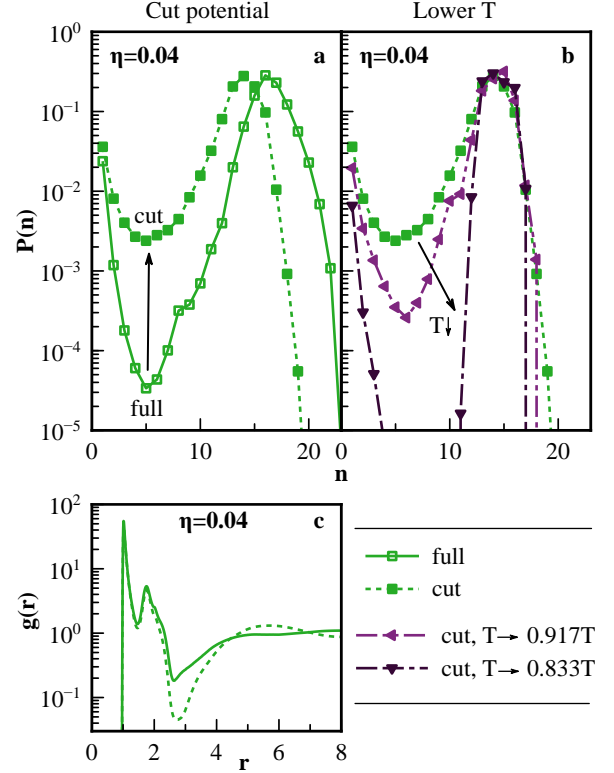}
  \caption{(a) and (c) Effect of cutting and shifting one \(n_{\text{tgt}}=16\) optimized potential (at the first minimum beyond the primary repulsion) on the CSD and RDF respectively (b) Evolution of the cut potential CSD from panel (a) with modest temperature rescaling.}
  \label{sch:FigureCUTANDCOOL}
\end{figure}

In Fig.~\ref{sch:FigureCUTANDCOOL}(a), we compare the CSDs corresponding to both the fully optimized and the truncated potential for \(\eta=0.04\): both potentials result in a clustered system that is fluid at both the intra- and inter-cluster level, indicating that the longer-ranged tail is not a strict requirement for IC-like behavior. However, the truncated potential yields clusters that are, on average, smaller (\(n^{*}=14\)) than \(n_{\text{tgt}}\) as well as less size-specific, as evident from the nearly two orders of magnitude increase in the CSD minumum between the \(n=1\) and primary \(n^{*}\) peaks. Nonetheless, more ideal clustering behavior can be restored by decreasing temperature by less than \(0.2k_{\text{B}}T\), as shown in Fig.~\ref{sch:FigureCUTANDCOOL}(b). These changes with \(T\) can be understood in the context of shifting the microscale liquid-gas co-existence towards the liquid side (i.e., favoring clusters over monomers) upon cooling. Overall, these observations support the notion that while the longer-ranged tail modulates the cluster-to-monomer ratio (and thus the monodispersity of the aggregates), it is not required to form ICs; thus, cluster formation in our systems is predominantly a result of the competitive broad attraction well and repulsive hump.

\begin{figure}
  \includegraphics{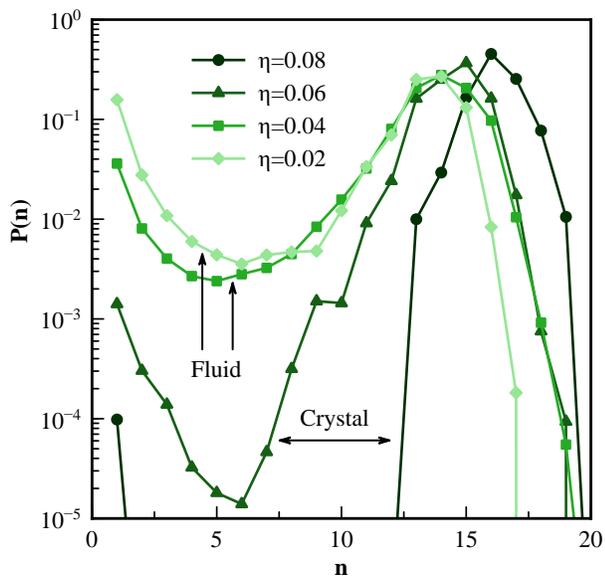}
  \caption{Effect of cutting and shifting the \(n_{\text{tgt}}=16\) optimized potentials on the CSD for all volume fractions.}
  \label{sch:FigureCUTCRYSTALLIZE}
\end{figure}

The RDFs for the full and truncated potentials are compared in Fig. ~\ref{sch:FigureCUTANDCOOL}(c). The obvious depression of the intercluster depletion region suggests that upon truncation, we effectively increase the repulsive footprint of a cluster. As a result, intercluster crystallization will likely be harder to avoid. This is confirmed in Fig.~\ref{sch:FigureCUTCRYSTALLIZE} as the cut potentials associated with both of the higher density cases (\(\eta=0.06\) and 0.08) formed superlaticces.

Interestingly, the superlattice states have much better cluster definition and size preservation than the two lower density systems, \(\eta=0.02\) and 0.04, that remained as IC fluids. Whether the enhanced cluster preservation is a byproduct of crystallization or simply higher densities (less void space) is not entirely clear. However, we do note that, in general, crystallization in systems of predominantly repulsive particles requires a high level of monodispersity. This is something our emergent (not quenched in) cluster entities can adaptively realize in order to explore more thermodynamically desirable, crystalline regions of phase space. In stark contrast, the fluid state does not have any natural propensity towards well-defined, monodisperse clusters, (polydispersity and disorder is favored) thus making the design of an IC \emph{fluid} all the more compelling.

\section{Conclusions}

In this paper, we demonstrated a novel application of standard inverse design methodology, namely, the targeted fabrication of liquid state structure. This approach was successfully applied to discover a new class of pair potentials that stabilize ideal cluster (IC) fluids, comprised of long-lived, monodisperse, spherical fluid droplets with good center of mass mobility. As compared to equilibrium fluids of amorphous clusters generated via SALR potentials, the IC fluid states of the designed potentials displayed much greater size-specificity (i.e., more sharply peaked CSDs) with cluster sizes that were less sensitive to changes in overall density. Furthermore, using a new measure for cluster lifetime, the ICs were shown to persist longer than comparably sized amorphous SALR clusters and to maintain their identities on timescales relevant for cluster diffusion. 

The ability of the optimized potentials to stabilize ICs can be understood in terms of their general features: the broad (rather than short-range) attractive wells allow for many particles to closely pack before the relatively narrow repulsive barriers are sampled. Moreover, the repulsive barrier directly encodes the scale for aggregation (as evidenced by the proportionality of the barrier lengthscale \(r_{\text{rep}}\) and cluster radius \(\langle R \rangle\)) and furnishes the individual clusters with well-defined repulsive shells. By contrast, only from a Fourier-space perspective can a preferential length scale for ordering be gleaned from SALR potentials~\cite{PhysRevE.91.042312}. It seems intuitive that this disparity between the IBI-optimized and SALR potentials is responsible for the former's enhanced cluster size-specificity (even under compression). While there is a continuum of possible SALR functional forms that could in principle yield ideal clusters, our results suggest the opposite given that (1) the IBI scheme did not result in SALR potentials, and (2) no pairwise SALR fluids have been reported that display IC-like behavior for small cluster sizes (though low density systems of large clusters, albeit with significant free monomer, have been seen~\cite{LeoLue2014}).

In addition to their broad attractive basins and sharp repulsive barriers, the optimized potentials also possessed weak longer-ranged oscillatory tails; however, these tails were found to be non-essential for IC formation. Upon truncating the optimized potentials beyond the repulsive barrier, systems still displayed IC-like behavior, though with slightly reduced size-specificity (tending toward \(n^{*} < n_{\text{tgt}}\)). To intentionally steer the IBI optimization towards shorter-ranged potentials, it may be fruitful to modify the constrained MC simulations to reproduce the general features of the \(g(r)\) profiles associated with the truncated potentials. For instance, because truncating the potential enhanced the intercluster depletion region, one might choose (1) a stronger radius of gyration constraints to densify the clusters and/or (2) larger repulsive cluster shells to better separate the clusters. Optimizing for the resulting \(g_{\text{tgt}}(r)\) profiles may then naturally yield short-ranged potentials.

In closing, our success at optimizing for IC fluids demonstrates the impressive flexibility of pair potentials for generating intricate multiscale architectures. Even greater flexibility could be achieved by the inverse design of more complex patchy, or fully angularly dependent potentials; however, new schemes for solving the inverse statistical mechanics problem must be developed. An interesting application of such a method would be to extend our current results to a patchy particle model where the greater degrees of freedom, while adding complexity, would likely allow for practically simpler (though likely of similar spatial range) interactions.  More generally, liquid-state inverse design also opens the door to discovery of new kinetically arrested materials: given the propensity for the clusters to act as renormalized objects (e.g., COM crystallization), this could include glasses or gels of supraparticles created by quenching or compression out of the fluid state.

\section{Acknowledgments}
We thank Delia Milliron for helpful discussions during the preparation of this manuscript. This work was partially supported by the National Science Foundation (1247945), the Welch Foundation (F-1696), and the Gulf of Mexico Research Initiative. We acknowledge the Texas Advanced Computing Center (TACC) at The University of Texas at Austin for providing HPC resources.

\footnotesize{
%\bibliography{Bibliography} %your .bib file
%\bibliographystyle{rsc} %the RSC's .bst file
}

%%%%%%%%%%%%%%%%%%%%%%%%%%%%%%%%%%%%%%%%%%%%%
%%%%%%%%self contained bibliography%%%%%%%%%%
%%%%%%%%%%%%%%%%%%%%%%%%%%%%%%%%%%%%%%%%%%%%%

%merlin.mbs apsrev4-1.bst 2010-07-25 4.21a (PWD, AO, DPC) hacked
%Control: key (0)
%Control: author (8) initials jnrlst
%Control: editor formatted (1) identically to author
%Control: production of article title (-1) disabled
%Control: page (0) single
%Control: year (1) truncated
%Control: production of eprint (0) enabled
%

%%%%%%%%%%%%%%%%%%%%%%%%%%%%%%%%%%%%%%%%%%%%%
%%%%%%%%self contained bibliography%%%%%%%%%%
%%%%%%%%%%%%%%%%%%%%%%%%%%%%%%%%%%%%%%%%%%%%%

\end{document}

\renewcommand{\thefigure}{S\arabic{figure}}

%\preprint{APS/123-QED}

\title{Supplementary Material - Equilibrium cluster fluids: Pair interactions via inverse design}

\author{Ryan B. Jadrich}
%\email{rjadrich@utexas.edu}
\affiliation{McKetta Department of Chemical Engineering, University of Texas at Austin, Austin, Texas 78712, USA}

\author{Jonathan A. Bollinger}
%\email{jonathanabollinger@gmail.com}
\affiliation{McKetta Department of Chemical Engineering, University of Texas at Austin, Austin, Texas 78712, USA}

\author{Beth A. Lindquist}
%\email{kpj@che.utexas.edu}
\affiliation{McKetta Department of Chemical Engineering, University of Texas at Austin, Austin, Texas 78712, USA}

\author{Thomas M. Truskett}
\email{truskett@che.utexas.edu}
\affiliation{McKetta Department of Chemical Engineering, University of Texas at Austin, Austin, Texas 78712, USA}

\maketitle

\section{IBI potentials: all densities}
% ===============================================================================================================
% &&&&&
% &&&
% &

In Figs S1-S3 we provide the potentials optimized at each volume fraction studied in the main text (\(\eta=\)0.02, 0.04, 0.06, and the additional case of 0.08 for \(n_{\text{tgt}}=16\)). Evident from these figures is the very minor role that density plays in determining the optimized potentials. Structural features largely remain fixed spatially and only a weak depression of the amplitude is found with increasing density. The larger amplitudes found at the lower densities presumably reflect the greater entropic drive the potential is competing with to release particles into the void region between clusters (i.e., greater space).

\begin{figure}[htp]
  \includegraphics{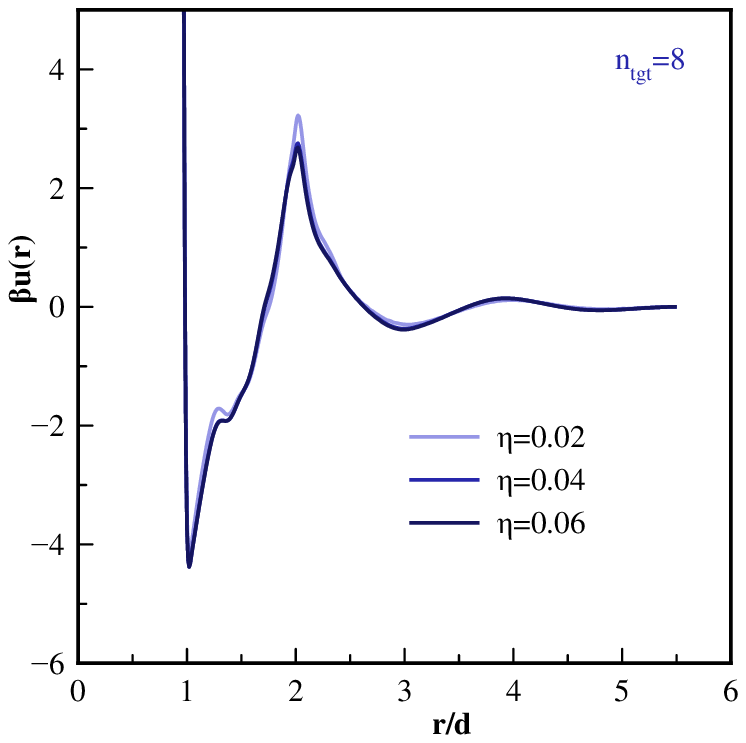}
  \caption{\linespread{1.0}\selectfont{} Optimized pair potentials for \(n_{\text{tgt}}=8\).}
  \label{sch:FigureS1}
\end{figure}

\begin{figure}[htp]
  \includegraphics{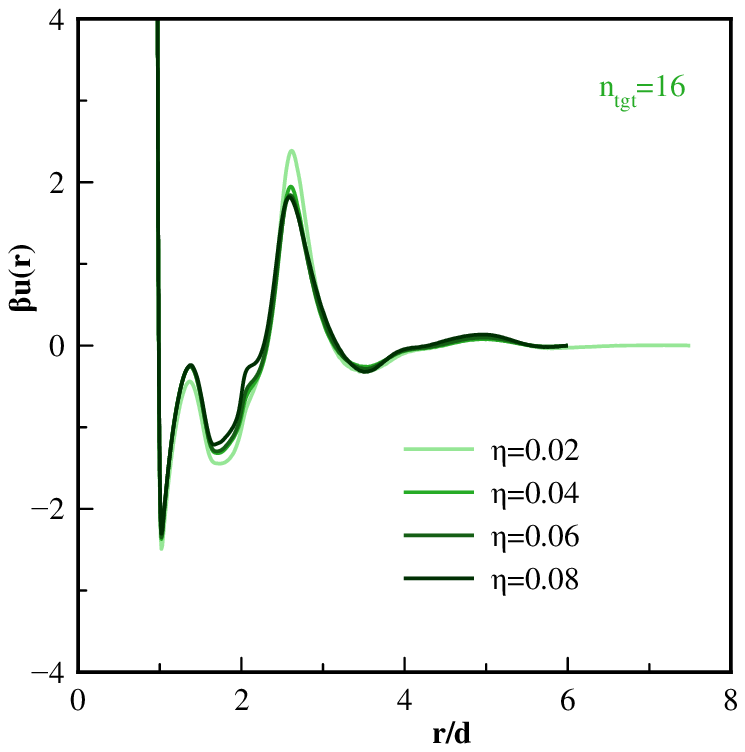}
  \caption{\linespread{1.0}\selectfont{} Optimized pair potentials for \(n_{\text{tgt}}=16\).}
  \label{sch:FigureS2}
\end{figure}

\begin{figure}[htp]
  \includegraphics{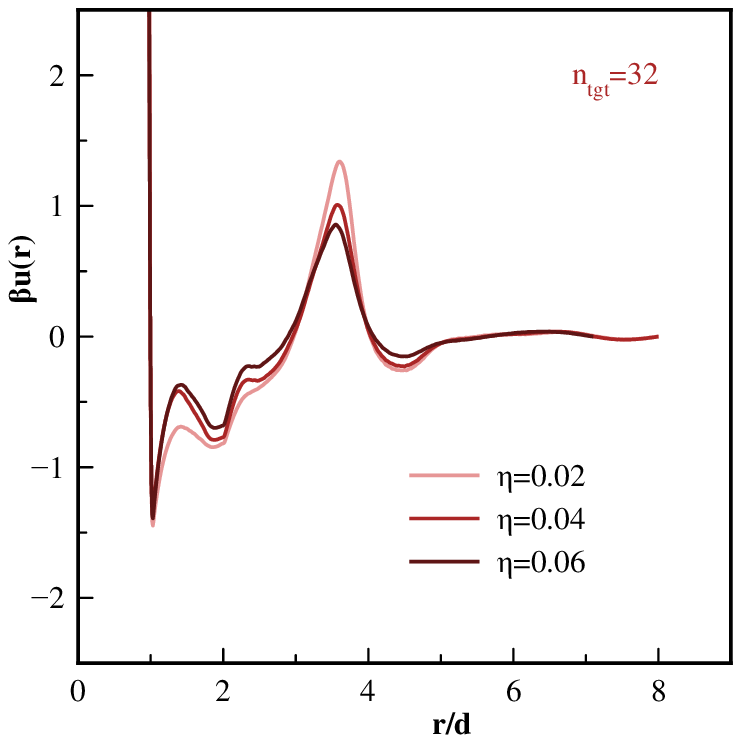}
  \caption{\linespread{1.0}\selectfont{} Optimized pair potentials for \(n_{\text{tgt}}=32\).}
  \label{sch:FigureS3}
\end{figure}

\section{Additional bond orientational order analysis data}

In Figs S4 and S5, we provide the bond-orientational order parameter distribution comparisons for \(n_{\text{tgt}}=8\) and \(16\), respectively. At a qualitative level, the agreement between the constrained and optimized systems is good; however, the agreement systematically decreases with decreasing cluster size as discussed in the main text.

\begin{figure}[htp]
  \includegraphics{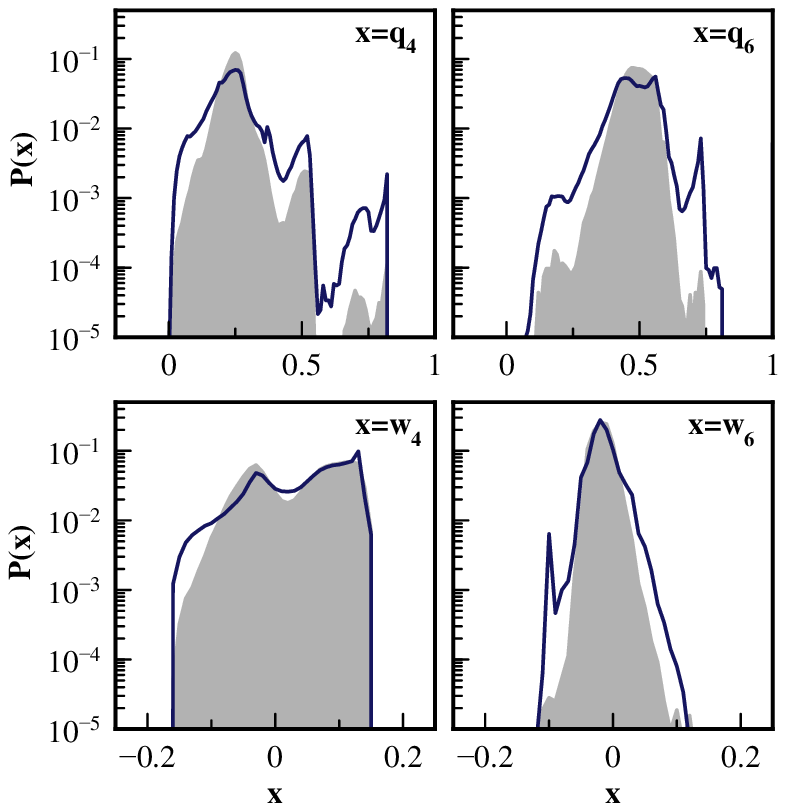}
  \caption{\linespread{1.0}\selectfont{} Bond-orientational order probability distributions for \(n_{\text{tgt}}=8\) at packing fraction \(\eta=0.06\). The filled grey and open blue curves indicate constrained and optimized simulations (using clusters with sizes between \(n^{*}\pm \delta n\) as provided in the main text), respectively.}
  \label{sch:FigureS4}
\end{figure}

\begin{figure}[htp]
  \includegraphics{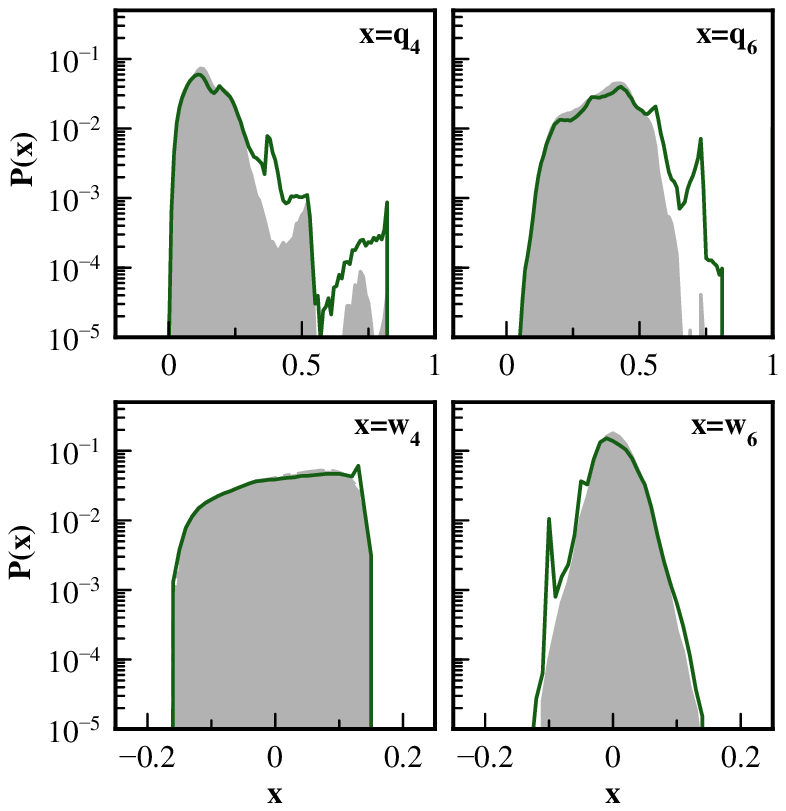}
  \caption{\linespread{1.0}\selectfont{} Bond-orientational order probability distributions for \(n_{\text{tgt}}=16\) at packing fraction \(\eta=0.06\). The filled grey and open green curves indicate constrained and optimized simulations (using clusters with sizes between \(n^{*}\pm \delta n\) as provided in the main text), respectively.}
  \label{sch:FigureS5}
\end{figure}

\newpage

%\bibliography{Bibliography}